# GRAMC: General-purpose and reconfigurable analog matrix computing architecture

Lunshuai Pan[1,2], Shiqing Wang[1,2], Pushen Zuo[1,2], Zhong Sun[1,2,3*]
[1]Institute for Artificial Intelligence, Peking University, Beijing, China
[2]School of Integrated Circuits, Peking University, China
[3]Beijing Advanced Innovation Center for Integrated Circuits
* Corresponding author: zhong.sun@pku.edu.cn

*Abstract*—In-memory analog matrix computing (AMC) with resistive random-access memory (RRAM) represents a highly promising solution that solves matrix problems in one step. However, the existing AMC circuits each have a specific connection topology to implement a single computing function, lack of the universality as a matrix processor. In this work, we design a reconfigurable AMC macro for general-purpose matrix computations, which is achieved by configuring proper connections between memory array and amplifier circuits. Based on this macro, we develop a hybrid system that incorporates an on-chip write-verify scheme and digital functional modules, to deliver a general-purpose AMC solver for various applications.

## I. Introduction

In recent years, in-memory AMC with nonvolatile resistive memory, for its high speed and low power consumption, has been extensively applied to accelerate matrix computations [1]. Nevertheless, the current in-memory AMC designs, such as those used in neural networks accelerators [2], linear system solvers [3],[4] and pseudoinverse computing unit [5], concerns only single computing functions, *e.g.*, matrix-vector multiplication (MVM), matrix inversion (INV), matrix pseudoinverse (PINV) or eigenvector (EGV), and hence they are inadequate to support general-purpose matrix computations.

These AMC solvers share same circuit components, *e.g.*, resistive memory array and analog amplifiers, but only with different connections of wiring these components. By reusing the circuit components, such a design also represents a strategy for saving hardware resources. In this work, we present a general-purpose and reconfigurable AMC architecture, termed GRAMC, for accelerating different kinds of matrix computations across multiple applications. It represents the first AMC architecture that supports both neural network accelerators and matrix equation solvers.

## II. Design of GRAMC

### A. On-Chip Write-Verify Scheme

The multi-level capability of RRAM devices is a great benefit for improving the equivalent throughput of AMC. To design an RRAM-based AMC macro, it is essential to develop an on-line write-verify scheme for multi-level RRAM programming. Fig. 1 illustrates the detailed write-verify scheme applied to a 1-transistor-1-resistor (1T1R) RRAM array. We adopt the Stanford-PKU RRAM open-source model [6], where the complex process of ion and vacancy immigration is simplified into the growth of a single domain filament that preserves the underlying physics (Fig. 1(a)). For one 1T1R cell, there are three terminals applied with voltages, including bit-line voltage ($V_{BL}$), source-line voltage ($V_{SL}$) and gate voltage ($V_g$), to control the write process [7]. During SET process, only $V_g$ is increased step by step, $V_{SL}$ is grounded and $V_{BL}$ is applied as $V_{set}$. By contrast, the RESET process is controlled by increasing $V_{SL}$. The conductance range of model is 1-100 μs, spanning from level 0 to level 15. The simulated data in Fig. 1(b), and 1(c) indicate that the write-verify scheme can be used to implement 16-level (4-bit) conductance states.

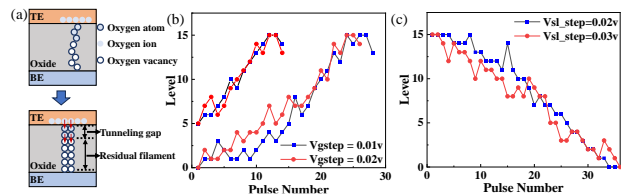

Fig. 1. (a) Schematic for filament growth. (b) Analog switching behaviors with different $V_g$ steps from different initial states in SET process, and (c) different $V_{SL}$ steps in RESET process. SET/RESET pulse width = 30 ns.

### B. AMC Macro Design

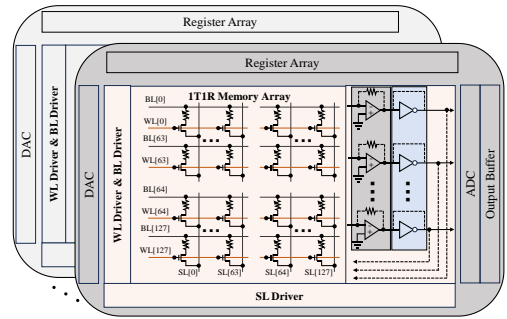

Fig. 2. An AMC macro group. Each AMC macro includes RRAM array, OPAs/TIAs, ADC, DAC, drivers, and output buffer. The register array controls the configuration of AMC circuits.

The AMC macro design is illustrated in Fig. 2. An AMC macro contains an RRAM array, and a set of operational amplifiers (OPAs) that can be reconfigured as TIAs and analog inverters. The size of RRAM array is moderately set as 128 × 128. The 1T1R cells in the crosspoint array are enabled by BL, WL, and source-line (SL) drivers, which allow to select the active region in the array to fit different sizes of matrix problems. The configuration messages are stored in the register array in advance and will control the transmission gates (on or off), thus configuring the connections between memory and OPAs. That said, the AMC macro can be reconfigured to perform different matrix computations. The DA/AD interfaces bridge the analog and digital domains, so that we can develop a hybrid design.

### C. System-level Design of GRAMC

The overall architecture of the GRAMC is shown in Fig. 3, which includes two main components: digital control module and a group of 16 AMC macros. The digital module is responsible for monitoring the program's running process and controlling the generation of messages.

The instructions from compiling stage will be loaded into the instruction stack in advance. Then, the instructions will be decoded to control the two data paths: write-verify path and system solution path. During the write-verify process (blue arrows), the voltages are applied to the memory array by DAC. Meanwhile, the output results by ADC will compare to the ideal values from global buffer in comparison units (CU). Until all the conductance states satisfy the error range or write pulse number is larger than the maximum pulse number, the write-verify process stops. Otherwise, the write-verify messages will be updated and the above process is repeated.

During the system solution process (red arrows), the detailed circuit connection messages from the decoder are stored in the register array. Subsequently, the external digital inputs from global buffer are converted to analog signals by DAC and then sent to the memory array to execute matrix computing. Finally, the computation results converted by ADCs will be stored in the output buffer. The output results may be further processed by digital functional modules for specific applications.

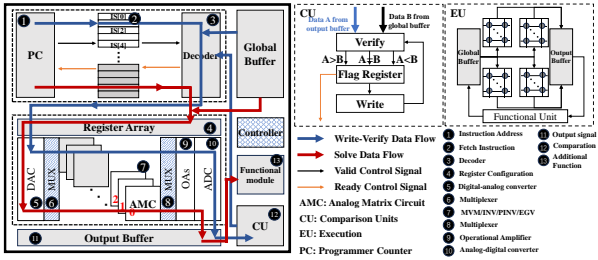

Fig. 3. The GRAMC architecture and data flow of write-verify (blue arrows) and system solution (red arrows).

## III. VALIDATION TEST

In this section, the GRAMC is reconfigured to execute the MVM, INV, PINV and EGV, respectively, thus demonstrating the reconfigurability. In circuit simulations, all matrices were mapped to one or two RRAM arrays with 4-bit quantization. The AMC results are compared to the numerical results from Python.

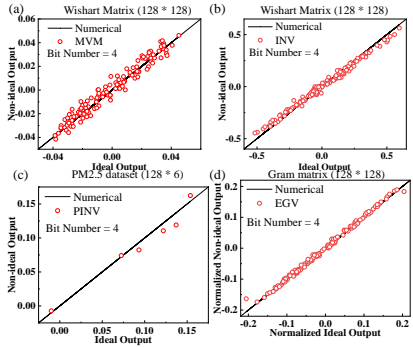

Fig. 4. Accuracy comparison between numerical and simulation results in GRAMC for (a) MVM, (b) INV, (c) PINV, and (d) EGV.

Fig. 4(a) and 4(b) show the results of MVM and INV of the $128 \times 128$ Wishart matrix, respectively. Fig. 4(c) shows the result from PINV circuit, which was reconfigured in the GRAMC to solve the linear regression task of $128 \times 6$. Fig. 4(d) shows the simulation result of EGV circuit reconfigured in GRAMC. The concerned matrix is a Gram matrix with size of $128 \times 128$. Generally, all circuits give approximate results compared to the numerical ones, showing relative errors around ten percent [4]. The computing accuracy is limited by the quantization error and the intrinsic analog noises in the circuit. Despite the deficiency of AMC results, they may be used as seed solutions to speed up the convergence towards precise final solutions.

As shown in Fig. 5, the LeNet5 is mapped in GRAMC to execute MNIST inference. The trained weights of each layer are loaded to the RRAM array by write-verify circuits. The convolutional computation results are transferred to the digital functional module to execute the pooling and activation operations. The test results show that with 4-bit quantization weights, GRAMC achieves a recognition accuracy of 97.1%. Bit slicing is a common strategy for improving computing accuracy of MVM [8]. Here, two RRAM arrays are used to store the most significant 4 bits and the least significant 4 bits of a weight matrix, respectively. As a result, the implemented LeNet-5 in GRAMC 8-bit weights improves the accuracy to 98.5%, which is only slightly lower than the accuracy (98.87%) with the float32 computing method.

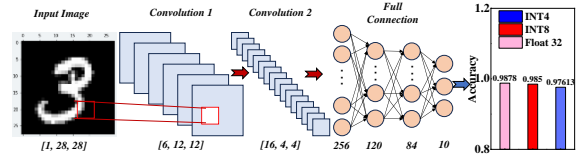

Fig. 5. Application of GRAMC to LeNet 5 neural network for MNIST dataset inference.

## IV. CONCLUSION

In this work, we propose the GRAMC, a general-purpose and reconfigurable analog matrix computing architecture that can be used to perform various matrix applications, ranging from MVM-based neural network accelerators to matrix equation solvers, including INV, PINV, and EGV. By combining these matrix primitives, including those nonlinear function units designed for neural networks, this system is applicable to more matrix problems. Such a capability is assisted by the design of on-chip write-verify scheme and digital functional modules in the system. While MVM-based neural network accelerator architectures have been intensively developed, the efforts in this work represent an important step towards efficient general-purpose analog matrix computers.

ACKNOWLEDGMENT

This work was supported by National Key R&D Program of China (No. 2020YFB2206001), National Natural Science Foundation of China (Nos. 92064004, 61927901), and the 111 project (No. B18001).